\documentclass[12pt]{article}
\usepackage[small]{titlesec}
\usepackage{amssymb,amsmath,amsthm}
\usepackage{enumerate}
\usepackage{hyperref}
\usepackage[mathscr]{euscript}
\usepackage[letterpaper,hmargin=3.7cm,vmargin=3.5cm]{geometry}

\usepackage{setspace}
\usepackage{graphicx}
\newtheorem{theorem}{Theorem}[section]

\newtheorem{lemma}[theorem]{Lemma}

\theoremstyle{definition}
\newtheorem{definition}[theorem]{Definition}
\newtheorem{remark}[theorem]{Remark}

\newcommand{\abs}[1]{\left\lvert #1 \right\rvert}
\newcommand{\norm}[1]{\left\lVert #1 \right\rVert}
\newcommand{\I}{i}
\newcommand{\inner}[2]{\left\langle#1,#2\right\rangle}

\newcommand{\cG}{\mathcal{G}}
\newcommand{\cH}{\mathcal{H}}
\newcommand{\cB}{\mathcal{B}}
\newcommand{\cL}{\mathcal{L}}
\newcommand{\cK}{\mathcal{K}}
\newcommand{\cP}{\mathcal{P}}
\newcommand{\cPW}{\mathcal{PW}}
\newcommand{\R}{{\mathbb R}}
\newcommand{\C}{{\mathbb C}}
\newcommand{\N}{{\mathbb N}}
\newcommand{\Z}{{\mathbb Z}}
\newcommand{\cc}[1]{\overline{#1}}
\newcommand{\ournewclass}{\mathscr{S}(\mathcal{H})}
\newcommand{\nentireclass}[1]{\mathscr{E}_{#1}(\mathcal{H})}
\newcommand{\graphel}[2]{\begin{pmatrix}#1\\ #2\end{pmatrix}}
\newcommand{\mb}{\boldsymbol}
\newcommand{\ds}{\displaystyle}

\renewcommand\tilde{\widetilde}
\renewcommand\hat{\widehat}

\def\cprime{$'$}

\DeclareMathOperator{\im}{im}
\DeclareMathOperator{\dom}{dom}

\DeclareMathOperator{\ran}{ran}
\DeclareMathOperator{\spec}{spec}
\DeclareMathOperator{\Span}{span}

\DeclareMathOperator{\assoc}{assoc}
\DeclareMathOperator{\mul}{mul}
\DeclareMathOperator{\Sp}{spec}
\DeclareMathOperator{\mt}{mt}

\begin{document}

\begin{titlepage}
\title
{\vspace{-1cm}De Branges spaces and Krein's theory of entire
operators
\footnotetext{%
Mathematics Subject Classification(2010):
46E22; 
Secondary
47A25, 
47B25, 
47N99. 
}
\footnotetext{%
Keywords: de Branges spaces, zero-free functions, entire operators.
}
\\[2mm]}
\author{
\textbf{Luis O. Silva}
\\
\small Departamento de F\'{i}sica Matem\'{a}tica\\[-1.6mm]
\small Instituto de Investigaciones en Matem\'{a}ticas Aplicadas y
	en Sistemas\\[-1.6mm]
\small Universidad Nacional Aut\'{o}noma de M\'{e}xico\\[-1.6mm]
\small C.P. 04510, M\'{e}xico D.F.\\[-1.6mm]
\small \texttt{silva@iimas.unam.mx}
\\[4mm]
\textbf{Julio H. Toloza}\thanks{Partially supported by CONICET (Argentina)
	through grant PIP 112-201101-00245.}
\\
\small CONICET\\[-1.6mm]
\small Centro de Investigaci\'{o}n en Inform\'{a}tica para la
	Ingenier\'{i}a\\[-1.6mm]
\small Universidad Tecnol\'{o}gica Nacional --
	 Facultad Regional C\'{o}rdoba\\[-1.6mm]
\small Maestro L\'{o}pez esq.\ Cruz Roja Argentina\\[-1.6mm]
\small X5016ZAA C\'{o}rdoba, Argentina\\[-1.6mm]
\small \texttt{jtoloza@scdt.frc.utn.edu.ar}}

\date{}
\maketitle
\begin{center}
\begin{minipage}{5in}
  \centerline{{\bf Abstract}} \bigskip
  This work presents a contemporary treatment of Krein's entire
  operators with deficiency indices $(1,1)$ and de Branges' Hilbert
  spaces of entire functions. Each of these theories played a central
  role in the research of both renown mathematicians.  Remarkably,
  entire operators and de Branges spaces are intimately connected and
  the interplay between them has had an impact in both spectral
  theory and the theory of functions. This work exhibits the
  interrelation between Krein's and de Branges' theories by means of a
  functional model and discusses recent developments, giving
  illustrations of the main objects and applications to the spectral
  theory of difference and differential operators.
\end{minipage}
\end{center}
\bigskip
\thispagestyle{empty}
\end{titlepage}
\tableofcontents
\section{Introduction}
\label{sec:introduction}

In a series of papers \cite{MR0012177,MR0011170,MR0011533} M. G. Krein
formulated the foundations of the theory of entire operators that
systematized some abstract essential facts shared by various,
seemingly unrelated, classical problems of mathematical analysis such
as the moment problem, the continuation of positive definite
functions, and the theory of spiral curves in Hilbert spaces. This
unifying approach eventually allowed to tackle other problems in
various fields and revealed interesting connections between
them. Krein's main motivation for constructing the theory of entire
operators seems to have been the classical moment problem since he
considered the works on the matter to be the germ of the theory (see
\cite{MR0011170} and \cite[Appendix 3]{MR1466698}). Entire operators
were present (not always explicitly) in a large part of Krein's
mathematical research and they occupied a prominent position in his
panoramic lectures at the jubilee session of the Moscow Mathematical
Society (1964) \cite[Appendix 3]{MR1466698} and the International
Congress of Mathematicians (1966) \cite{MR0235424}.


Krein's theory of entire operators combines methods of operator
theory, particularly spectral theory, and the theory of analytical
functions, particularly entire functions. This combination has
produced an interplay of ideas between these two fields that has been
very fruitful in both areas. Here, it is pertinent to mention that
Krein developed new results and posed new problems in the theory of
functions because of his investigations related to entire
operators.

The connection of operator theory and the theory of functions
mentioned above arises from the modeling of a symmetric operator of a
certain class as the operator of multiplication by the independent
variable in a certain functional space. This key part of the theory of
entire operators was called by Krein the representation theory of
symmetric operators, but it is actually a functional model (see
Section~\ref{sec:funct-model}). Functional models for various classes
of operators have been studied by various authors throughout the
history of operator theory. The best known functional model is the
so-called canonical form of a simple selfadjoint operator
\cite[Section 69]{MR1255973}. This model is obtained via the spectral
theorem. Other instances are functional models for contractions
\cite{MR0275190} and dissipative operators \cite{MR0365199,MR0385642}
(see also \cite{MR0500225,MR513172}). It is worth remarking that
Krein's approach to the construction of functional models for
symmetric operators was further developed and generalized by Strauss
in his theory of functional models for closed linear operators
\cite{MR1660000,MR1759535,MR1821925}.

For any entire operator with deficiency indices (1,1), Krein's
functional model yields a Hilbert space of scalar entire
functions. Krein noticed that this space had very distinctive
properties \cite[Appendix 3]{MR1466698} and studied some of them in
the course of his research on the theory of entire operators. The
Hilbert spaces of entire functions corresponding to entire operators
were the first instances of the spaces that were later introduced and
studied by L. de Branges who was not aware of Krein's results. The
works by de Branges on the theory of Hilbert spaces of entire
functions
\cite{debranges0,debranges1,debranges3.5,debranges2,debranges3,debranges4}
(which were later compiled in the book \cite{debranges}), were
considered very deep and far reaching by Krein \cite[Appendix
3]{MR1466698}.

The theory of Hilbert spaces of entire functions has played a central
role in de Branges research work. This theory is an important
ingredient in his celebrated proof of the Bieberbach conjecture
\cite{debranges6}.  Noteworthily, de Branges theory has been applied
to various aspects of spectral theory of differential operators
\cite{dym,eckhardt-1,remling-1,us-6}.

Krein not only studied entire operators with deficiency indices
$(1,1)$. He also incursioned into the investigation of entire
operators with finite and infinite deficiency indices. Krein's
functional model in the case of arbitrary finite and equal deficiency
indices yields to Hilbert spaces of vector entire
functions. Coincidentally, de Branges studied also spaces of vector
entire functions \cite{MR0244795}, however these spaces are no longer,
strictly speaking, de Branges spaces and complications arise when
dealing with the parallels between the theory of these spaces and
Krein's entire operators. For this reason, since this review paper
deals with the relations between de Branges spaces and Krein's entire
operators, the discussion is restricted to the case of deficiency
indices $(1,1)$.

This work is not exhaustive, many things were deliberately left out in
order to keep the material neat, handy, and user-friendly.  The aim of
this review paper is to introduce the reader to this theory which has
multiple ramifications and is interconnected with many objects in
analysis and analytic function theory.


\section{On a class of symmetric operators}
\label{sec:ourclass}
This section introduces the class of symmetric operators relevant to
this work and recollects material on operator and spectral theories
that will be used in the course of the exposition. A more refined
classification of symmetric operators, containing the main object of
the present chapter, is given by the end of the section.

\subsection{Closed symmetric operators and their selfadjoint extensions}
\label{sec:symmetric-operators}

Let $\cH$ be a complex Hilbert space with inner product $\inner{\cdot}{\cdot}$,
the latter assumed antilinear in its first argument.
To any linear operator $T$ acting within $\cH$, there
corresponds a linear subset
\begin{equation*}
  \left\{\graphel{\phi}{\psi}\in
\mathcal{H}\oplus\mathcal{H}: \phi\in\dom(T),\ \psi=T\phi\right\}
\end{equation*}
which is called its graph. In this section it is useful to identify an
operator with its graph. By this approach an operator is a particular
case of a linear relation which is, by definition, a linear subset of
$\mathcal{H}\oplus\mathcal{H}$. In this work, all operators and
relations are linear.

For any relation $T$, one has
\begin{equation}
  \label{eq:def-kmdr}
 \begin{split}
  \ker(T):=\left\{\phi\in\cH:\graphel{\phi}{0}\in T\right\},&\quad
  \mul(T):=\left\{\phi\in\cH:\graphel{0}{\phi}\in T\right\},\\
  \dom(T):=\left\{\phi\in\cH:\graphel{\phi}{\psi}\in T\right\},&\quad
  \ran(T):=\left\{\psi\in\cH:\graphel{\phi}{\psi}\in T\right\}.
\end{split}
\end{equation}
The relation $T$ is an operator if and only if $\mul(T)=\{0\}$. A
relation is closed if it is a closed set with respect to the norm in
$\cH\oplus\cH$, that is, a closed relation is a subspace of
$\cH\oplus\cH$. Thus, an operator is closed if and only if its graph
is a subspace of $\cH\oplus\cH$.

For any operator $T$, its adjoint $T^*$ is defined by
\begin{equation}
\label{eq:adjoint-def}
 T^*:= \left\{\graphel{\eta}{\omega}\in\cH\oplus\cH :
		\inner{\eta}{T\phi}=\inner{\omega}{\phi}
		\mbox{ for all }\phi\in\dom(T)\right\},
\end{equation}
where $T^*$ is an operator whenever (\ref{eq:adjoint-def}) is the
graph of an operator and a multivalued relation otherwise. It is
straightforward to verify that $T^*$ is an operator if and only if
$\dom(T)$ is dense in $\cH$ \cite[Lemma 3, Section 1, Chapter 3]{MR1192782}.

Let $A$ be a closed linear operator which is symmetric, that is,
$A\subset A^*$ (as subsets of $\cH\oplus\cH$). It is also assumed that
the deficiency indices
\begin{align*}
  n_+(A):&=\dim[\cH\ominus\ran(A-zI)],\quad z\in\C^+,\\
  n_-(A):&=\dim[\cH\ominus\ran(A-zI)],\quad z\in\C^-.
\end{align*}
are such that
\begin{equation}
  \label{eq:deficiency-indices-assumption}
  n_+(A)=n_-(A)=1\,.
\end{equation}
Since the operator $A$ is closed and symmetric, $\ran(A-zI)$ is
closed whenever $\im(z)\ne 0$. Thus, for any nonreal $z$ the Hilbert
space $\cH$ admits the decomposition into subspaces
\begin{equation}
  \label{eq:space-decomposition}
  \cH=\ran(A-zI)\oplus\ker(A^*-\cc{z}I)
\end{equation}
(see \cite[Theorem~5, Section 3, Chapter 3]{MR1192782} for the case when
$\cc{\dom(A)}=\cH$ and \cite[Proposition~3.31]{MR0123188} for the
general case). Now, in view of (\ref{eq:space-decomposition}), the
assumption on the deficiency indices
(\ref{eq:deficiency-indices-assumption}) implies that
\begin{equation}
  \label{eq:dim-ker-1}
  \dim\ker(A^*-zI)=1\quad\text{for all}\quad z\in\C\setminus\R\,.
\end{equation}
According to (\ref{eq:def-kmdr}) and (\ref{eq:adjoint-def}), one has
\begin{equation*}
 \mul(A^*) = \left\{\omega\in\cH : \inner{\omega}{\psi} = 0
		\mbox{ for all }\psi\in\dom(A)\right\}.
\end{equation*}
Therefore $\mul(A^*) = \dom(A)^\perp$.

Besides the symmetric operator $A$, this work deals with its
canonical selfadjoint extensions.  A canonical selfadjoint extension of
a given symmetric operator is a selfadjoint extension within the
original space $\cH$. In other words, a canonical selfadjoint
extension $A_\gamma$ of $A$ satisfies
\begin{equation*}
  A\subset A_\gamma=A_\gamma^*\subset A^*,\quad\text{as subsets of
  }\  \cH\oplus\cH.
\end{equation*}
Since the restriction of an operator is an operator, one obviously has
that all canonical selfadjoint extensions of $A$ are operators
whenever $\cc{\dom(A)}=\cH$. If $A$ is nondensely defined, this is no
longer true. However, under the condition imposed on the deficiency
indices (\ref{eq:deficiency-indices-assumption}), the situation is not
quite dissimilar.
\begin{theorem}
\label{thm:misc-about-symm-operators}
Let $A$ be a closed, nondensely defined, symmetric operator in a
Hilbert space.
If (\ref{eq:deficiency-indices-assumption}) holds,  then:
\begin{enumerate}[{(i)}]
\item The codimension of $\dom(A)$ equals one.
\item All except one of the canonical selfadjoint extensions of $A$
	are operators.
\end{enumerate}
\end{theorem}
A proof of this theorem follows
from \cite[Section 1, Lemma 2.2 and Theorem 2.4]{MR1430397} (see
also \cite[Proposition 5.4]{MR1430397} and the comment below it). This
work deals only with canonical selfadjoint extensions.

The spectral properties of the selfadjoint extensions of $A$ are
essential in this work and, in view of
(ii) above, the
reader is reminded that the spectrum of a closed linear relation $T$ in $\cH$,
denoted $\Sp(T)$ is the complement of the set of all $z\in\C$ such
that $(T-zI)^{-1}$ is a bounded operator defined on all $\cH$.
Moreover, $\Sp(T)\subset\R$ when $T$ is a selfadjoint linear relation
\cite[Theorem 3.20]{MR0361889}.

\subsection{Generalized Cayley transform}
\label{sec:gen-cayley-transform}
A closed symmetric operator $A$ with equal finite deficiency indices has always
canonical selfadjoint extensions, with some of them being proper linear
relations if $\dom(A)$ is not dense in $\cH$ 
(Theorem~\ref{thm:misc-about-symm-operators} describes this fact when the
both deficiency indices are equal to one). In any case, the resolvent
of a given canonical selfadjoint extension, say $A_\gamma$, is always an operator.
Given such an extension of $A$, define
\begin{equation}
  \label{eq:generalized-cayley}
  	V_\gamma(w,z):=(A_\gamma -wI)(A_\gamma-zI)^{-1}=
        I + (z-w)(A_\gamma-zI)^{-1},
\end{equation}
for $w\in\C$ and $z\in\C\setminus\spec(A_\gamma)$.  This operator is
the generalized Cayley transform of $A_\gamma$. Unlike the (regular) Cayley
transform of a selfadjoint operator, $V_\gamma(w,z)$ is not unitary
for arbitrary values of $w$ and $z$ where it is defined. This
operator, however, has various relevant properties. Indeed, directly from the
first resolvent identity \cite[Equation 12, Section 7, Chapter 3]{MR1192782} 
(which also holds when $A_\gamma$ is a relation), one verifies that for any 
$v,w,z\in\C\setminus\spec(A_\gamma)$
\begin{equation}
  \label{eq:gen-cayley-properties1}
  V_\gamma(w,z)=V_\gamma(z,w)^{-1},\qquad
  V_\gamma(w,z)V_\gamma(z,v)=V_\gamma(w,v).
\end{equation}
Also, it is straightforward to establish that
\begin{equation}
  \label{eq:gen-cayley-props2}
  V_\gamma(w,z)^*=V_\gamma(\cc{w},\cc{z}).
\end{equation}
By means of the first identity in (\ref{eq:gen-cayley-properties1})
and (\ref{eq:gen-cayley-props2}), the following simple assertion is
proven.
\begin{theorem}
  \label{thm:misc-about-symm-operator2}
  For any choice of a canonical selfadjoint extension $A_\gamma$ of a closed
  symmetric operator $A$, the operator $V_\gamma(w,z)$ maps
  $\ker(A^*-wI)$ injectively onto $\ker(A^*-zI)$.
\end{theorem}

Define the function
\begin{equation}\label{eq:mapping-in-kernel}
\psi(z):= V_\gamma(w_0,z)\psi_{w_0},
\end{equation}
for given $\psi_{w_0}\in\ker(A^*-w_0I)$ with $w_0\in\C$. It follows
from Theorem~\ref{thm:misc-about-symm-operator2} that $\psi(z)$ is in
$\ker(A^*-z I)$.  Clearly, $\psi(z)$ is an analytic function in
$\C\setminus\spec(A_\gamma)$ because of the analytic properties of the
resolvent.  Obviously, $\psi(w_0)=\psi_{w_0}$. Moreover, as a
consequence of the second identity in
(\ref{eq:gen-cayley-properties1}), one has
\begin{equation}\label{eq:identity-between-kernels}
\psi(z) = V_\gamma(v,z)\psi(v),
\end{equation}
for any pair $z,v\in\C\setminus\spec(A_\gamma)$.

Note that, in this subsection,
(\ref{eq:deficiency-indices-assumption}) is not relevant. All
assertions on the properties of the generalized Cayley transform only
require the existence of a selfadjoint extension of $A$, i. e., the
equality of the deficiency indices.

\subsection{Complete nonselfadjointness}
\label{sec:complete-nonselfadjointness}

\begin{definition}
\label{def:complete-nonselfadjointness}
A closed symmetric nonselfadjoint operator is said to be completely
nonselfadjoint if it is not a nontrivial orthogonal sum of a symmetric
and a selfadjoint operators.
\end{definition}

 Since an invariant subspace of a
symmetric operator is a subspace reducing that operator
\cite[Theorem 4.6.1]{MR1192782}, a symmetric operator $A$ is completely
nonselfadjoint when there is not a nontrivial invariant subspace of $A$ on
which $A$ is selfadjoint.

With the help of Theorem \ref{thm:misc-about-symm-operator2} it
can be proven that
\begin{equation*}
  \bigcap_{z\in\C\setminus\R}\ran(A-zI)
\end{equation*}
is the maximal invariant subspace in which $A$ is selfadjoint. Hence,
a necessary and sufficient condition for the symmetric operator $A$ to
be completely nonselfadjoint is
\begin{equation}
  \label{eq:simplicity}
  \bigcap_{z\in\C\setminus\R}\ran(A-zI)=\{0\}
\end{equation}
(see \cite[Proposition 1.1]{MR0463964} for the general case and
\cite[Theorem 1.2.1]{MR1466698} for the densely defined case).
Note that, due to (\ref{eq:space-decomposition}), the condition
(\ref{eq:simplicity}) is equivalent to
\begin{equation}
  \label{eq:simplicity-equivalent}
  \cc{\Span_{z\in\C\setminus\R}\{\ker(A^*-zI)\}}=\cH\,.
\end{equation}

Complete nonselfadjointness plays an important role in this work's
further considerations. Here, some of the distinctive features that a
closed symmetric operator has when it is completely nonselfadjoint are
briefly discussed. Consider the function $\psi(z)$ given by
\eqref{eq:mapping-in-kernel} and take a sequence
$\{z_k\}_{k=1}^\infty$ with elements in
$\mathbb{C}\setminus\mathbb{R}$ having accumulation points in the
upper and lower half-planes. Suppose that there is $\eta\in\cH$ such
that $\inner{\eta}{\psi(z_k)}=0$ for all $k\in\mathbb{N}$. This
implies that $\inner{\eta}{\psi(z)}=0$ for
$z\in\mathbb{C}\setminus\mathbb{R}$ because of the analyticity of the
function $z\mapsto\inner{\eta}{\psi(z)}$. Therefore, by
\eqref{eq:simplicity-equivalent}, one concludes that $\eta=0$. Thus,
completely nonselfadjoint, closed symmetric operators can exist only
in a {\em separable} Hilbert space. From now on, the reader should
assume that $\cH$ is separable.

As in the previous subsection, the condition
(\ref{eq:deficiency-indices-assumption}) was so far not assumed in
the ongoing discussion. However, for the next property related to complete
nonselfadjointness, it is required that
(\ref{eq:deficiency-indices-assumption}) holds. First, some definitions:

\begin{definition}
  \label{def:involution}
 A mapping $J$ of $\cH$ onto itself such that, for any
 $\phi,\psi\in\cH$ and $a,b\in\C$,
 \begin{equation*}
   J(a\phi+b\psi)=\cc{a}J\phi+\cc{b}J\psi,\quad
   J^2=I,\quad\text{and}\quad\inner{J\psi}{J\phi}=\inner{\phi}{\psi},
 \end{equation*}
is called an involution.
\end{definition}
\begin{definition}
  \label{def:commutativity-of-invoution}
An involution $J$ is said to commute with a selfadjoint relation $T$
if
\[
J(T-zI)^{-1}\phi = (T-\cc{z}I)^{-1}J\phi,
\]
for every $\phi\in\cH$ and $z\in\C\setminus\R$. If $T$ is moreover
an operator this is equivalent to the usual notion of commutativity,
that is,
\[
J\dom(T)\subseteq\dom(T),\qquad JT\phi = TJ\phi,
\]
for every $\phi\in\dom(T)$.
\end{definition}
\begin{theorem}
\label{thm:existence-of-commuting-involution}
  Let $A$ be a completely nonselfadjoint, closed, symmetric operator with
  deficiency indices $n_+(A)=n_-(A)=1$. Then there exists an involution $J$
  that commutes with all its canonical selfadjoint extensions.
\end{theorem}

The proof of this assertion is constructive. As was already shown, the
complete
nonselfadjointness of $A$ implies that the sequence
$\{\psi(z_k)\}_{k=1}^\infty$, used in the paragraph following
(\ref{eq:simplicity-equivalent}), is total. Define
\[
J\left(\sum_{k=1}^Nc_k\psi(z_k)\right): = \sum_{k=1}^N\cc{c_k}\psi(\cc{z_k})\,,
\]
for some $N\in\N$. Then $J$ is extended to the whole space and has the
property
\begin{equation*}
  J\psi(z)=\psi(\cc{z})\,,\quad \text{for all}\ z\in\C\setminus\spec(A_\gamma)\,.
\end{equation*}
Using the properties (\ref{eq:gen-cayley-properties1}) and
(\ref{eq:gen-cayley-props2}) of the generalized Cayley transform, it
is shown that $J$ is an involution and commutes with
$A_\gamma$. Finally, due to (\ref{eq:deficiency-indices-assumption}),
a generalization of Krein's resolvent formula (see \cite[Theorem
3.2]{MR1430397}) implies the result (some details of this proof can be
found in \cite[Proposition~2.3]{us-4}).

The following assertion (cf. \cite[Proposition~2.11]{us-4}) is related
to the previous one and, again, its proof relies on the assumption
(\ref{eq:deficiency-indices-assumption}).

\begin{theorem}
\label{thm:on-psi-w}
Let $A$ be a completely nonselfadjoint, closed, symmetric operator
with deficiency indices $n_+(A)=n_-(A)=1$, and $J$ be an involution
that commutes with a canonical selfadjoint extension $A_\gamma$ of $A$
(hence it commutes with all canonical selfadjoint
extensions). For every $v\in\Sp(A_\gamma)$, there exists
$\psi_v\in\ker(A^*-vI)$ such that $J\psi_v=\psi_v$.
\end{theorem}

\subsection{Regularity}
\label{sec:regularity}

\begin{definition}
\label{def:regularity}
A closed operator $T$ is regular if for every $z\in\C$
there exists $c_z>0$ such that
\begin{equation}
\label{eq:regular-point}
\norm{(T-zI)\phi}\ge c_z\norm{\phi},
\end{equation}
for all $\phi\in\dom(T)$.
\end{definition}

In other words, $T$ is regular when every
point of the complex plane is a point of regular type.

\begin{remark}
\label{rem:regularity-implies-simplicity}
It is easy to see that a regular, closed symmetric operator is
necessarily completely nonselfadjoint as regularity
implies that the spectral kernel is empty and, therefore, the operator cannot
have selfadjoint parts. On the other hand, there are completely
nonselfadjoint operators that are not regular.
\end{remark}

Since the residual spectrum of a regular operator $T$ fills up the whole complex
plane, it follows from \cite[Section 7.3 Chapter 3]{MR1192782} that every
complex number is an eigenvalue of $T^*$.

Consider again the case of a closed symmetric operator $A$
with equal deficiency indices and assume that $A$ is regular. If
$A_\gamma$ is any canonical selfadjoint extension of $A$, then, since
$A_\gamma$ is a restriction of $A^*$, it follows from
\cite[Section 7.4, Chapter 3]{MR1192782} that 
$\spec(A_\gamma)=\spec_\text{pp}(A_\gamma)$,
that is, every element of the spectrum is an eigenvalue.

The following theorem is well known for the case when the
operator is densely defined. The proof in the general case can
be found in \cite[Proposition 2.4]{us-4}.

\begin{theorem}
\label{thm:properties-of-new-class}
Let $A$ be a regular, closed, symmetric operator such that
(\ref{eq:deficiency-indices-assumption}) holds. The following
assertions are true:
\begin{enumerate}[{(i)}]
\item The spectrum of every canonical selfadjoint extension of $A$
	consists solely of isolated eigenvalues of multiplicity one.
\item Every real number is part of the spectrum of one, and only one,
	canonical selfadjoint extension of $A$.
\item The spectra of the canonical selfadjoint extensions of $A$ are
	pairwise interlaced.
\end{enumerate}
\end{theorem}

Note that (i) above implies that every selfadjoint extension of $A$ is
a simple operator \cite[Section 69]{MR1255973}.

\subsection[Classes of symmetric operators]
			{The classes $\mb{\ournewclass}$ and $\mb{\nentireclass{n}}$}
\label{sec:classes-of-symmetric-operators}

In this subsection, the main classes of operators
considered in this work are introduced.

\begin{definition}
  \label{def:ourclass}
 The class $\ournewclass$ is the set of all regular, closed symmetric
 operators with both deficiency indices equal 1, that is,
 \begin{equation*}
   \ournewclass:=\{A\text{ is a regular, closed symmetric operator}:
   	n_+(A)=n_-(A)=1\}.
 \end{equation*}
\end{definition}

By Remark~\ref{rem:regularity-implies-simplicity}, all operators in
$\ournewclass$ are completely nonselfadjoint. Furthermore, for any
element of $\ournewclass$, Theorem~\ref{thm:properties-of-new-class}
holds, and, by Theorem~\ref{thm:existence-of-commuting-involution},
one can construct an involution that commutes with all its selfadjoint
extensions.

\begin{definition}
  \label{def:n-entireclass+}
  An operator $A\in\ournewclass$ is said to belong to the class
  $\nentireclass{n}$, $n\in\N\cup\{0\}=\Z^+$, if there exist $n+1$
  vectors $\mu_0,\dots,\mu_n\in\cH$ such that
\begin{equation}
  \label{eq:def-En-classes}
	\cH = \ran(A-zI)\dotplus\Span\{\mu_0+z\mu_1+\cdots +z^n\mu_n\},
        \quad\text{for all }z\in\C\,.
\end{equation}
\end{definition}
The class $\nentireclass{0}$ admits a further breaking up into the
following subclasses.
\begin{definition}
  \label{def:n-entireclass-}
 An operator $A\in\nentireclass{0}$ is in $\nentireclass{-n}$, $n\in\N$, if
there exists a vector $\mu_{-n}\in\dom(A^{n})$ such that
\begin{equation*}
	\cH = \ran(A-zI)\dotplus\Span\{\mu_{-n}\},
	\quad\text{for all }z\in\C\,.
\end{equation*}
\end{definition}
Thus, there is an operator class $\nentireclass{n}$ for any
$n\in\Z$. Moreover, one has the following chain of inclusions
\begin{equation*}
  \dots\subset\nentireclass{-1}\subset\nentireclass{0}\subset\nentireclass{1}
  \subset\dots\subset\ournewclass.
\end{equation*}
The following notation will be used
\begin{equation}
  \label{eq:e-infinity}
  \nentireclass{-\infty}:=\bigcap_{n\in\Z}\nentireclass{n}.
\end{equation}
It turns out that the class $\nentireclass{-\infty}$ is the class of
nonselfadjoint Jacobi operators (see \S~\ref{sec:hamburger}). Also it
is easy to see that \cite[Example 3.13]{us-4}
\begin{equation*}
  \bigcup_{n\in\Z^+}\nentireclass{n}\subsetneq\ournewclass\,.
\end{equation*}

An operator of the class $\nentireclass{n}$, $n\in\Z$, will be
henceforth called $n$-entire. The classes $\nentireclass{0}$ and
$\nentireclass{1}$ correspond to those defined originally by Krein;
this point will be elucidated later in
\S~\ref{sec:kreins-entire-operators}.

\section{On de Branges Hilbert spaces}

Most of the elementary albeit profound aspects of the theory of de
Branges Hilbert spaces were introduced by L. de Branges himself in
\cite{debranges0,debranges1,debranges3.5,debranges2,debranges3,debranges4},
and later compiled and given some further development in
\cite{debranges}. An introductory and more amenable exposition of this
theory, intended toward its application to the spectral analysis of
Sturm-Liouville operators, can be found in \cite{dym}. Another
introductory presentation is found in \cite[Chapter 6]{dym-mckean}. In
passing, it is worth mentioning that the de Branges Hilbert space
theory has been generalized to Pontryagin spaces of entire functions,
an ambitious task being carried out by M. Kaltenb{\"a}ck and
H. Woracek in a series of papers \cite{kaltenback-1,kaltenback-2,
  kaltenback-3,kaltenback-4,kaltenback-6,kaltenback-5,woracek-1}.

\subsection{Definition and elementary properties}
There are two essentially different ways of defining a de Branges
Hilbert space (dB space from now on). The one introduced next is
of axiomatic nature.

\begin{definition}
\label{def:axiomatic-db}
A Hilbert space of entire functions $\cB$ is an (axiomatic) dB space if and only
if, for every function $f(z)$ in $\cB$, the following conditions holds:
\begin{enumerate}[({A}1)]
\item For every $w\in\C$, the linear functional
        $f(\cdot)\mapsto f(w)$  is continuous;

\item for every non-real zero $w$ of $f(z)$, the function
        $f(z)(z-\cc{w})(z-w)^{-1}$ belongs to $\cB$
        and has the same norm as $f(z)$;

\item the function $f^\#(z):=\cc{f(\cc{z})}$ also belongs to $\cB$
        and has the same norm as $f(z)$.
\end{enumerate}
\end{definition}

By Riesz lemma, condition (A1) is equivalent to saying that $\cB$ has a
reproducing kernel, that is, there exists a function $k:\C\times\C\to\C$
such that, for every $w\in\C$, the function $z\mapsto k(z,w)$ belongs to
$\cB$ and has the property
\[
\inner{k(\cdot,w)}{f(\cdot)}_{\cB}=f(w)\text{ for all }
		f(z)\in\cB;
\]
here $\inner{\cdot}{\cdot}_{\cB}$ denotes the inner product in $\cB$ (assumed
linear in the second argument). Moreover,
\[
k(w,w)=\inner{k(\cdot,w)}{k(\cdot,w)}_\cB\ge 0
\]
where, as a consequence of (A2), the positivity is strict for every non-real $w$
as long as $\cB$ contains a nonzero element
\cite[Lemma 1]{debranges0}. Note that
\[
k(z,w)=\inner{k(\cdot,z)}{k(\cdot,w)}_\cB,
\]
therefore $k(w,z)=\cc{k(z,w)}$. Furthermore, (A3) implies that
$\cc{k(\cc{z},w)}\in\cB$ for every $w\in\C$ from which it can be shown that
$\cc{k(\cc{z},w)}=k(z,\cc{w})$ \cite[Lemma 1]{debranges0}.
Since by the previous discussion one obtains $k(w,z)=k(\cc{z},\cc{w})$,
it follows that $k(z,w)$ is anti-entire with respect to the second argument
(it is obviously entire with respect to the first one).

The other way of defining a dB space is constructive and
requires two ingredients. The first one is the Hardy space
\[
\cH_2(\C^+):=\left\{f(z)\text{ holomorphic in }\C^+:
		\sup_{y>0}\int_\R\abs{f(x+\I y)}^2 dx<\infty\right\},
\]
where $\C^+:=\{z=x+\I y: y>0\}$. The second ingredient is an Hermite-Biehler
function (HB function for short), that is, an entire function $e(z)$
such that $\abs{e(z)}>\abs{e(\cc{z})}$ for all $z\in\C^+$.

\begin{definition}
\label{def:canonical-db}
The (canonical) dB space associated with an HB function $e(z)$ is the
linear manifold
\[
\cB(e):=\left\{f(z)\text{ entire}:
        \frac{f(z)}{e(z)},\frac{f^\#(z)}{e(z)}\in\cH_2(\C^+)\right\},
\]
equipped with the inner product
\[
\inner{f(x)}{g(x)}_{\cB(e)}:=\int_{R}\frac{\cc{f(x)}g(x)}{\abs{e(x)}^2} dx.
\]
\end{definition}
Thus defined, $\cB(e)$ is a Hilbert space \cite[Theorem
21]{debranges}, indeed, it can trivially be identified with a subspace
of $L^2(\R,\abs{e(x)}^{-2}dx)$.

The set $\cB(e)$ can be characterized as
\begin{equation}
\label{eq:db-space-alternate}
\cB(e)=\left\{\begin{gathered}
		f(z)\text{ entire}:
 		\int_{R}\abs{\frac{f(x)}{e(x)}}^2 dx<\infty\text{ and }
 		\\[1mm]
        \abs{\frac{f(z)}{e(z)}}\le\frac{c_f}{\sqrt{\im(z)}},\,
        \abs{\frac{f^\#(z)}{e(z)}}\le\frac{c_f}{\sqrt{\im(z)}}
        \text{ for all }z\in\C^+
        \end{gathered}\right\}
\end{equation}
\cite[Proposition 2.1]{remling-1} so one can alternatively define $\cB(e)$ by
(\ref{eq:db-space-alternate}).

Definitions \ref{def:axiomatic-db} and \ref{def:canonical-db} are equivalent
(as expected) in the following sense \cite[Problem 50 and Theorem 23]{debranges};
see also \cite[Section 6.1]{dym-mckean}.

\begin{theorem}
\label{thm:equiv-db-spaces}
Let $\cB$ be an axiomatic dB space that contains a nonzero element. Then
there exists an HB function $e(z)$ such that $\cB=\cB(e)$
isometrically. Conversely, for every HB function $e(z)$, the
associated canonical dB space $\cB(e)$ satisfies (A1), (A2) and (A3).
\end{theorem}

Given an HB function $e(z)$, the reproducing kernel of $\cB(e)$ can be written
as \cite[Theorem~19]{debranges}
\[
k(z,w) = \begin{cases}
		 \dfrac{e^\#(z)e(\cc{w}) - e(z)e^\#(\cc{w})}{2\pi\I(z-\cc{w})},
					&w\ne\cc{z},
		 \\[1mm]
		 \frac{1}{2\pi\I}\left[e^{\#'}(z)e(z) - e'(z)e^\#(z)\right],
					&w=\cc{z}.
		 \end{cases}
\]
On the other hand, for a given dB space $\cB$ an HB function that
makes Theorem \ref{thm:equiv-db-spaces} hold is
\[
e(z) = \I\sqrt{\frac{\pi}{\im(w_0)k(w_0,w_0)}}(z-\cc{w_0})k(z,w_0),
\]
where $w_0$ is some fixed number in $\C^+$ \cite[Lemma 4]{debranges0}.
Note that a given dB space $\cB$ is not associated to a unique
HB function, as it is apparent from the previous formula. However, the
different HB functions that give rise to the same dB space are all related
in the precise form asserted below \cite[Theorem 1]{debranges1}.
The following statement makes use of the customary decomposition 
$e(z) = a(z) - \I b(z)$, where
\[
a(z) := \frac{e(z)+e^\#(z)}{2},\quad
b(z) := \I\frac{e(z)-e^\#(z)}{2}.
\]
Notice that these newly introduced entire functions are real in the sense that
they satisfy the identity $f^\#(z)=f(z)$.

\begin{theorem}
Suppose $M$ is a $2\times 2$ real matrix such that $\det M=1$.
Let $e(z)=a(z)-\I b(z)$ be an HB function. Define $e_M(z):=a_M(z)-\I b_M(z)$,
where
\[
\begin{bmatrix}
a_M(z) \\ b_M(z)
\end{bmatrix}
= M \begin{bmatrix}
	a(z) \\ b(z)
	\end{bmatrix}.
\]
Then $e_M(z)$ is an HB function and $\cB(e_M)=\cB(e)$ isometrically. Conversely,
if $\tilde{e}(z)$ is an HB function such that $\cB(\tilde{e})=\cB(e)$
isometrically, then $\tilde{e}(z)=e_M(z)$ for some $2\times 2$ real matrix $M$.
\end{theorem}


Orthogonal sets in a dB space can be constructed by mean of phase
functions \cite[Theorem 22]{debranges}. A phase function
associated with an HB function $e(z)$ is a real, monotonically
increasing continuous (indeed differentiable) function $\phi(x)$ such
that $e(x)\exp[i\phi(x)]\in\R$ for all $x\in\R$ \cite[Problem
48]{debranges}.

\begin{theorem}
Given $\cB=\cB(e)$, let $\phi(x)$ be a phase function associated with $e(z)$.
Then, for every $\alpha\in\R$, the following assertions hold true:
\begin{enumerate}[(i)]
\item $\cK_\alpha:=\left\{\frac{k(z,t_n)}{e^\#(t_n)}:
		t_n\in\R\text{ such that }\phi(t_n)=\alpha\!\!\mod\pi\right\}$ is
		an orthogonal set in $\cB$;
\item $\Span\cK_\alpha\ne\cB(e)$ if and only if
		$e^{i\alpha}e(z)-e^{-i\alpha}e^\#(z)\in\cB$;
\item if $e^{i\alpha}e(z)-e^{-i\alpha}e^\#(z)\not\in\cB$ then
		\[
		\norm{f(\cdot)}_\cB^2
			= \sum_n\abs{\frac{f(t_n)}{e(t_n)}}^2\frac{\pi}{\phi'(t_n)},
		\]
		for every $f(z)\in\cB$.
\end{enumerate}
\end{theorem}

In connection with the last theorem it is worth mentioning that,
associated with every orthogonal set $\cK_\alpha$ of a dB space $\cB$
(and assuming $\Span\cK_\alpha=\cB$), one has the sampling formula
\[
f(z)=\sum_n\frac{k(z,t_n)}{k(t_n,t_n)}f(t_n)
	=\sum_n\frac{g(z)}{(z-t_n)g'(t_n)}f(t_n),\quad f(z)\in\cB,
\] 
where the latter expression has the form of a standard Lagrange interpolation
formula; here $g(z)=\frac{1}{2\pi\I}\left[e^\#(z)e(t_n)-e(z)e^\#(t_n)\right]$.

\subsection{Spaces of associated functions}

\begin{definition}
An entire function $h(z)$ is said to be associated to a dB space $\cB$ if
\[
\frac{f(z)h(w)-f(w)h(z)}{z-w}\in\cB,
\]
for every $f(z)\in\cB$ and $w\in\C$ such that $f(w)\ne 0$. The set of
all functions associated to $\cB$ is denoted by  $\assoc\cB$.
\end{definition}

Clearly, $\assoc\cB$ is a linear manifold which can also be constructed
in terms of $\cB$ itself \cite[Lemma 4.5]{kaltenback-1},
\[
\assoc\cB = \cB + z\cB.
\]
Another insightful characterization of $\assoc\cB$ is given by
\cite[Theorem 25]{debranges}, which can be formulated as follows
\cite[p. 236]{kaltenback-a}.

\begin{theorem}
Let $e(z)$ be an HB function such that $\cB=\cB(e)$. Then
\[
\assoc\cB = \left\{f(z)\text{ \rm entire}:
			 \frac{f(z)}{(z+i)e(z)},\frac{f^\#(z)}{(z+i)e(z)}\in\cH_2(\C^+)
			 \right\}.
\]
\end{theorem}
Note that the characterization above implies that $\assoc\cB(e)$ itself
turns out to be
a dB space, since $(z+i)e(z)$ is an HB function. Furthermore, it is also clear
that $e(z)\in\assoc\cB(e)\setminus\cB(e)$.

Within $\assoc\cB(e)$ there is a distinguished family of functions, defined
as
\[
s_\beta(z):=\frac{\I}{2}\left[e^{\I\beta}e(z)-e^{-\I\beta}e^\#(z)\right],
	\quad \beta\in[0,\pi).
\]
Generically $s_\beta(z)\in\assoc\cB(e)\setminus\cB(e)$. More precisely
\cite[Lemma 7]{debranges0}:

\begin{lemma}
Assume $\cB(e)$ contains a nonzero element. Then at most one of the
functions $s_\beta(z)$ belongs to $\cB(e)$.
\end{lemma}

A special role is played by the zeros of the functions $s_\beta(z)$.

\begin{lemma}
Let $e(z)$ be an HB function having no real zeros. Assume furthermore that
$\cB(e)$ contains a nonzero element. Then, for every $\beta\in[0,\pi)$,
the zeros of $s_\beta(z)$ are all simple and real. Moreover, the zeros of any
two functions $s_\beta(z)$ and $s_\gamma(z)$, with $\beta\ne\gamma$, are
interlaced.
\end{lemma}

The notion of functions associated to a dB space has been generalized
in \cite{langer-1,woracek-1}:

\begin{definition}
\label{def:associated-functions}
Given $n\in\Z$, the set of $n$-associated functions of a dB space $\cB$
is
\[
\assoc_n\cB :=\begin{cases}
				\cB + z\cB + \cdots + z^n\cB, & n \ge 0,
				\\
				\dom(S^{\abs{n}}), & n<0.
			  \end{cases}
\]
\end{definition}

These linear sets also become dB spaces when equipped with suitable inner
products; see \cite[Corollary 3.4]{langer-1} and \cite[Example 2.7]{woracek-1}.

An important result concerns the existence of a real zero-free
function $n$-associated to a dB space; see \cite[Theorem
5.1]{langer-1}, \cite[Theorem 3.2]{woracek-1}, and \cite[Theorem
2.7]{us-6} for further details.

\begin{theorem}
\label{thm:1-in-assoc-n-boosted}
  Suppose $e(x)\neq 0$ for all $x\in\mathbb{R}$ and
  $e(0)=(\sin\gamma)^{-1}$ for some fixed $\gamma\in(0,\pi)$. Furthermore
  assume that $\dim\cB(e)=\infty$. Let
  $\{x_j\}_{j\in\mathbb{N}}$ be the sequence of zeros of the function
  $s_\gamma(z)$. Also, let $\{x_j^+\}_{n\in\mathbb{N}}$ and
  $\{x_j^-\}_{n\in\mathbb{N}}$ be the sequences of positive,
  respectively negative, zeros of $s_\gamma(z)$, arranged according
  to increasing modulus.  Then a zero-free, real entire function
  belongs to $\assoc_n\mathcal{B}(e)$ if and only if the following
  conditions hold true:
\begin{enumerate}[(C1)]
\item The limit
	$\displaystyle{\lim_{r\to\infty}\sum_{0<|x_j|\le r}
		\frac{1}{x_j}}$
	exists.
\item $\displaystyle{\lim_{j\to\infty}\frac{j}{x_j^{+}}
		=- \lim_{j\to\infty}\frac{j}{x_j^{-}}<\infty}$.
\item Assuming that $\{b_j\}_{n\in\mathbb{N}}$ are the zeros of
  $s_\beta(z)$, define
	\[
	h_\beta(z):=\left\{\begin{array}{ll}
			\displaystyle{\lim_{r\to\infty}\prod_{|b_j|\le r}
			\left(1-\frac{z}{b_j}\right)}
				& \mbox{ if 0 is not a root of } s_\beta(z),
			\\
			\displaystyle{z\lim_{r\to\infty}\prod_{0<|b_j|\le r}
			\left(1-\frac{z}{b_j}\right)}
				& \mbox{ otherwise. }
			   \end{array}\right.
	\]
	The series
	$\displaystyle{
		\sum_{x_j\ne 0}\abs{\frac{1}
		{x_j^{2n}h_{0}(x_j)h_{\gamma}'(x_j)}}}$ is convergent.
\end{enumerate}
\end{theorem}
\begin{remark}
  \label{rem:zero-free-implies-real-zero-free}
  In the previous theorem, the assumption that a real zero-free function 
  exists in $\cB$ can be weakened to just requiring the existence of a zero-free
  function (not necessarily real) in $\cB$. This follows from the fact that 
  if a zero-free function is in $\cB$ then a real zero-free function is also in
  $\cB$ \cite[Remark 2.8]{us-4}.
\end{remark}

\subsection{The operator of multiplication by the independent variable}
\label{sec:multiplication-operator-in-dB}
Given a dB space $\cB$, the operator of multiplication by the independent
variable is defined by
\begin{gather*}
\dom(S) = \{f(z)\in\cB: zf(z)\in\cB\};\quad
(Sf)(z)=zf(z),\quad f(z)\in\dom(S).
\end{gather*}

The basic properties of this operator are summarized next.

\begin{theorem}
\label{thm:multiplication-operator}
Let $S$ be the operator defined as above in a dB space $\cB$. Then:
\begin{enumerate}[(i)]
\item $S$ is closed, symmetric, completely nonselfadjoint, and has
		deficiency indices $(1,1)$;

\item $S$ is real with respect to the involution $\#$, that is,
		$f^\#(z)\in\dom(S)$ whenever $f(z)\in\dom(S)$ and $(Sf)^\#(z)=zf^\#(z)$;

\item $\dom(S)$ is dense in $\cB$ if and only if $s_\beta(z)\not\in\cB$
		for all $\beta\in[0,\pi)$;

\item if otherwise $s_\gamma(z)\in\cB$ for some (necessarily unique)
		$\gamma\in[0,\pi)$, then $\dom(S)^\perp=\Span\{s_\gamma(z)\}$.
\end{enumerate}
Assume moreover that $\cB=\cB(e)$ isometrically for some HB function $e(z)$
having no real zeros. Then:
\begin{enumerate}[(i)]
\setcounter{enumi}{4}
\item $S$ is regular;

\item $\ran(S-wI)^\perp=\Span\{k(z,w)\}$, for every $w\in\C$.
\end{enumerate}
\end{theorem}

In view of (iii) and (iv) above, all the canonical selfadjoint extensions
of $S$, with at most the exception of one, are operators. In any case their
resolvents can be described as follows
\cite[Propositions 4.6 and 6.1]{kaltenback-1}:

\begin{theorem}
The canonical selfadjoint extensions of $S$ have a bijective correspondence
with the associated functions $s_\beta(z)$, $\beta\in[0,\pi)$. This
correspondence is given by
\begin{equation}
\label{eq:seladjoint-resolvents}
(S_\beta-wI)^{-1}f(z) := \frac{f(z)-\frac{s_\beta(z)}{s_\beta(w)}f(w)}{z-w},
\end{equation}
for every $f(z)\in\cB(e)$ and
$w\in\C\setminus\{\text{\rm zeros of }s_\beta(z)\}$. Moreover, $S_\beta$ is
a proper linear relation if an only if $s_\beta(z)\in\cB(e)$, in which case
$\mul S_\beta = \Span\{s_\beta(z)\}$.
\end{theorem}

Whenever $s_\beta(z)\not\in\cB(e)$ the resolvent description
\eqref{eq:seladjoint-resolvents} is equivalent to the following one,
\begin{gather}
\dom(S_\beta)  :=
	\left\{g(z) = \frac{s_\beta(w)f(z)-s_\beta(z)f(w)}{z-w},\,
		   f(z)\in\cB(e),\, w\in\C\right\},
	\label{eq:domain-selfajoint-s}
	\\[1mm]
(S_\beta g)(z) :=
	z g(z) + f(w)s_\beta(z).
	\nonumber
\end{gather}
This later form can be obtained from the standard von Neumann theory when
$S$ has dense domain in $\cB(e)$ \cite[p. 367]{us-2}. In this case the
adjoint operator can likewise be specified
in terms of the function $e(z)$. Namely,
\begin{gather*}
\dom(S^*)  :=
	\left\{\begin{gathered}
	g(z)=\frac{f(z)-\frac{e(z)}{e(w)}f(w)}{z-w} +
	            \frac{h(z)-\frac{e^\#(z)}{e^\#(\cc{w})}h(\cc{w})}{z-\cc{w}},\\
		   f(z),h(z)\in\cB(e),\, w:\im w> 0
		   \end{gathered}\right\},
    \\[1mm]
(S^*g)(z) :=
	z g(z) + \frac{e(z)}{e(w)}f(w) + \frac{e^\#(z)}{e^\#(\cc{w})}h(\cc{w}).
\end{gather*}

When $\dom(S)$ is not dense in $\cB(e)$, the canonical selfadjoint operator
extensions of $S$ can be written also as a family of rank-one perturbations
\cite[Theorem 3.4 and 3.5]{us-5}:

\begin{theorem}
Assume $s_0(z)\in\cB$. Then the set of canonical selfadjoint operator
extensions of $S$ is given by
\begin{equation*}
\dom(S_\beta)=\dom(S_{\pi/2}),\qquad S_\beta =
S_{\pi/2}-\frac{\cot\beta}{\pi}\inner{s_0(\cdot)}{\cdot\,}_\cB s_0(z),
\end{equation*}
for $\beta\in(0,\pi)$. Moreover, for every $\beta\in(0,\pi)$,
$s_0(z)$ is a generating vector for the operator $S_{\beta}$.
\end{theorem}

From \eqref{eq:domain-selfajoint-s} it is clear that the eigenvectors of
the selfadjoint operator extensions are also related
to the functions $s_\beta(z)$.
Up to a normalization they are given by $\frac{s_\beta(z)}{z-x^{\beta}_n}$,
where $x^{\beta}_n\in\spec(S_\beta)$ is the associated eigenvalue (hence
$s_\beta(x^{\beta}_n)=0$).

\subsection{Isometric inclusion of dB spaces}

A subspace $\cL$ of a dB space $\cB$ is a dB subspace if itself is a
dB space with respect to the inner product inherited from $\cB$.

A distinctive structural property of dB spaces states that the set of
dB subspaces of a given dB space is totally ordered by inclusion
\cite[Theorem 35]{debranges}. Below is a simplified version of this
assertion, taken from \cite[Section 6.5]{dym-mckean}. In what follows
a dB space $\cB$ is called regular (or short, according to
\cite{dym-mckean}) if $1\in\assoc\cB$.

\begin{theorem}
  Assume $\cB_1$, $\cB_2$ and $\cB_3$ are regular dB spaces such that
  $\cB_1$ and $\cB_2$ are both isometrically contained in
  $\cB_3$. Then either $\cB_1\subset\cB_2$ or $\cB_2\subset\cB_1$.
\end{theorem}

Recall that the mean type of a function $f(z)$ (of bounded type on $\C^+$;
for a definition of this see \cite[Section 2]{kaltenback-a}) is the number
\[
\mt f := \limsup_{y\to+\infty}\frac{1}{y}\log\abs{f(\I y)}.
\]
Note that, for an HB function $e(z)$, the mean type of $e^\#(z)/e(z)$ is
always nonpositive (this follows from the very definition of an HB function).

The following is an abridged version of \cite[Theorem
2.7]{kaltenback-a}. Notice that here no condition of regularity is
assumed.

\begin{theorem}
Given a dB space $\cB(e)$, define $\tau_e = \mt\frac{e^\#}{e}$
and $e_\tau(z) = e(z)e^{\I \frac{\tau}{2}z}$. Then:
\begin{enumerate}[(i)]
\item $e_\tau(z)$ is an HB function if and only if $\tau\ge\tau_e$;
\item for each $\tau\in[\tau_e,0]$, $\cB(e_\tau)$ is a dB subspace of
		$\cB(e)$;
\item the chain of all dB subspaces $\cB(\tilde{e})$ of $\cB(e)$, where
		$\tilde{e}(z)$ has the same real zeros as $e(z)$ (counting
		multiplicities), is given by
		$\left\{\cB(e_\tau):\tau\in[\tau_e,0]\right\}$.
\end{enumerate}
\end{theorem}

\subsection{Examples}
\label{sec:examples-db-spaces}

A rather comprehensive discussion of special classes of dB spaces is
found in \cite[Chapter 3]{debranges}. Here, some more or less
ubiquitous examples are presented.

\paragraph{Spaces of polynomials.} Consider the linear set
\[
\cP_n:=
	\{\text{polynomials of degree} < n\}.
\]
The set $\cP_n$ can be turned out into a dB space by choosing an HB
function from $\assoc\cP_n\setminus\cP_n=\cP_{n+1}\setminus\cP_n$
(here $\assoc\cP_n=\cP_n+z\cP_n$). That is (allowing some abuse of
notation), $\cP_n=\cB(p_n)$ where $p_n(z)$ is a polynomial of degree
$n$ with all its roots in $\C^-$. Clearly, $\cP_n$ has finite
dimension, $\dim\cP_n=n$. Conversely, if $\cB=\cB(e)$ is a dB space of
dimension $\dim\cB=m<\infty$ then there exists a zero-free, real
entire function $g(z)$ such that $e(z)=g(z)p_{m}(z)$, where $p_{m}(z)$
is a polynomial of degree $m$ whose roots are in $\C^-$ \cite[Problem
88]{debranges}.  In other words, $\cB=g(z)\cP_{m}$ as sets.

The operator of multiplication $S$ is not densely defined in $\cP_n$,
in fact, $\dom(S)=\cP_{n-1}$. Also,
$\assoc_k\cP_n=\cP_{n+k}$. Finally, one verifies that the chain of dB
subspaces of $\cP_n$ is $\{\cP_m:m\le n\}$.

It is worth mentioning at this point that a subject of great interest
in the theory of dB spaces concerns the role of polynomials. Contrary
to a somewhat naive intuition may suggest, an infinite dimensional dB
space may has no polynomials at all, it may contain polynomials only
up to certain finite degree, or even if all of them are included in
the space they may not be a dense subset. In connection with the
latter case, note that if the polynomials are contained in a dB
space $\cB$ (assumed infinite dimensional), then they are all in the
domain of the operator $S$. Thus a trivial necessary condition for the
set of polynomials to be dense in $\cB$ is that $S$ be densely
defined. Also note that if $1\in\assoc\cB\setminus\cB$, then no
polynomial belongs to $\cB$; conspicuous examples are the Paley-Wiener
spaces introduced below. There is a good deal of research done on this
topic, see for instance \cite{baranov} (and references therein).

\paragraph{Paley-Wiener spaces.}

Arguable, they are the prototypical, as well as motivational, examples
of dB spaces, as attested by de Branges himself \cite{debranges}. The
Paley-Wiener space, parametrized by $a>0$, is defined as
\[
\cPW_a:=\left\{f(z)\text{ entire}: \int_\R\abs{f(x)}^2dx<\infty\text{
    and } \abs{f(z)}\le c_fe^{a\abs{z}}\right\},
\]
equipped with the inner product of $L^2(\R)$. It can be shown that
\[
\cPW_a =\left\{f(z)\text{ entire}: \int_\R\abs{f(x)}^2dx<\infty\text{
    and } \abs{f(z)}\le
  c'_f\frac{e^{a\abs{\im(z)}}}{\abs{\im(z)}^{1/2}}\right\}
\]
(see for instance \cite[Chapter 6]{dym-mckean}). In view of
\eqref{eq:db-space-alternate} one obtains $\cPW_a=\cB(e^{-\I az})$;
note that $e^{-\I az}$ is an HB function as long as $a\ge 0$. The
Paley-Wiener theorem states that every function in $\cPW_a$ is the
analytic continuation (to the entire complex plane) of the Fourier
transform of a function in $L^2(-a,a)$. That is,
\[
\cPW_a=\left\{f(z)\text{ entire}: f(z)=\int_{-a}^ae^{-\I zx}\varphi(x) dx,\quad
		       \varphi(x)\in L^2(-a,a)\right\}.
\]

No polynomial belongs to $\cPW_a$. However, it is easy to verify that
if $p_n(z)$ is a polynomial of degree $n\ge 0$, then
$p_n(z)\in\assoc_{n+1}\cPW_a\setminus\assoc_{n}\cPW_a$. Also, the
chain of dB subspaces of $\cPW_a$ is $\{\cPW_b:b\in(0,a]\}$.

Paley-Wiener spaces have many more distinctive properties; further details
are accounted for in \cite[Chapter 2]{debranges}.

\paragraph{dB spaces associated to Bessel functions.} This kind of dB
spaces appears in connection with the radial Hamiltonian operator of a
quantum free particle in spherical coordinates \cite[Section 3]{us-6};
see also \S \ref{sec:radial-schrodinger-operator} below.

Given $l\in\Z^+$ and $b>0$, define
\[
\cG^l_b := \left\{\begin{gathered}
				   f(z)\text{ entire}: f(z)=f(-z), 
			       \int_0^b x^{2l+2}\abs{f(x)}^2dx<\infty,
			       \\
			       \abs{z^{l+1}f(z)}\le c_fe^{b\abs{\im(z)}}\text{ for all }z\in\C
			       \end{gathered}\right\}
\]
equipped with the inner product of $L^2(\R^+;x^{2l+2}dx)$. By a
theorem due to Griffith \cite{griffith} (see also \cite{zemanian}),
which is to some extend a generalization of the Paley-Wiener theorem but
involving the Hankel transform, one has
\[
\cG^l_b = \left\{f(z)\text{ entire}:
			z^{l+1}f(z) = \int_0^b \sqrt{zx}J_{l+\frac12}(zx)\varphi(x)dx,
			\quad \varphi(x)\in L^2(0,b)\right\};
\]
here $J_m(w)$ denotes the Bessel function of order $m$. In order to verify
that $\cG^l_b$ is a dB space, define
\[
\xi_l(z,x):= z^{-(l+1)}\sqrt{zx}J_{l+\frac12}(zx).
\]
On one hand, in terms of this function, it holds true that
\[
\cG^l_b = \left\{f(z)\text{ entire}: f(z) = \int_0^b
  \xi_l(z,x)\varphi(x)dx, \quad \varphi(x)\in L^2(0,b)\right\}.
\]
On the other hand, $\xi_l(z,x)$ is the $L^2(0,b)$ fundamental solution
of the differential equation
\[
-\psi''(x) + \frac{l(l+1)}{x^2}\psi(x) = z^2\psi(x)
\]
with suitable boundary conditions at $x=0$ (for details, see
\cite[Section 3]{us-6}). An argument involving the Lagrange identity
(see \cite[Theorem 3.2]{eckhardt-1}) shows that $\cG^l_b=\cB(e^l_b)$,
where
\[
e^l_b(z) := \xi_l(z,b) + \I\xi'_l(z,b)
\]
(the prime denotes derivative with respect to the second argument).

These dB spaces do no contain polynomials. Moreover,
$1\in\assoc_{n_l+1}\cG^l_b\setminus\assoc_{n_l}\cG^l_b$, where
$n_l:=\lfloor\frac{l}{2}+\frac34\rfloor$ (the standard notation for
the floor function has been used here).

\section[A functional model for the operators discussed here]
		{A functional model for operators in $\mb{\ournewclass}$}
\label{sec:funct-model}

A functional model for a given operator $A$ in a Hilbert space $\cH$
is a unitary map of $\cH$ onto a Hilbert space $\widehat{\cH}$ of
\emph{functions} with certain analytical properties, such that the
operator $A$ is transformed into the operator of multiplication by the
independent variable in $\widehat{\cH}$.

This section describes a functional model for operators in
$\ournewclass$ which is suitable for studying the classes
$\nentireclass{n}$, $n\in\Z$. This functional model, which was
developed in \cite{us-2,us-4,us-3}, stems from Krein's theory of
representation of symmetric operators developed in his original work
\cite[Theorems 2 and 3]{MR0011170} (cf. \cite[Section
1.2]{MR1466698}), but differs from it in a crucial way as will be
explained below (see Remark~\ref{rem:differ-from-krein}).  The
functional model presented here can be viewed as a particular
realization (with some modifications) of the general theory developed
by Strauss \cite{MR1660000,MR1759535,MR1821925} and it is different
from (and simpler than) an equivalent functional model introduced in
\cite{MR2805419}.

\subsection{The functional space}
\label{sec:function-space}
Fix an operator $A\in\ournewclass$ and let $J$ be an involution that
commutes with the selfadjoint extensions of $A$. Consider a function
$\xi_A:\C\to\cH$ such that
\begin{enumerate}[({P}1)]
\item $\xi_A(z)$ is zero-free and entire,
\item $\xi_A(z)\in\ker(A^*-zI)$ for all $z\in\C$, and
\item $J\xi_A(z)=\xi_A(\cc{z})$ for every $z\in\C$.
\end{enumerate}

Since, for an operator $A\in\ournewclass$, one has that
$\dim\ker(A^*-zI)=1$ for all $z\in\C$,
the following assertion clearly holds true (see 
\cite[Proposition 2.12 and Remark 2.13]{us-4}).
\begin{lemma}
  \label{lem:xi-up-to-function}
  If $\xi_A^{(1)}:\C\to\cH$ and $\xi_A^{(2)}:\C\to\cH$ are two functions
  satisfying (P1),(P2), and (P3), then there exists a zero-free real
  entire function $g(z)$ such that $\xi_A^{(1)}(z)=g(z)\xi_A^{(2)}(z)$.
\end{lemma}

The function $\xi_A(z)$, which is crucial for the functional model
described below, can be constructed as follows. Pick a canonical
selfadjoint extension $A_\gamma$ of $A\in\ournewclass$ and let
$h_\gamma(z)$ be a real entire function whose zero set (counting
multiplicities) equals $\spec(A_\gamma)$ (hence, by (i) of
Theorem~\ref{thm:properties-of-new-class}, the zeros of $h_\gamma(z)$
are simple). On the basis of the analytical properties of the
generalized Cayley transform (see \S~\ref{sec:gen-cayley-transform})
and Theorem~\ref{thm:misc-about-symm-operator2}, it is straightforward
to verify that if one sets
\begin{equation}
  \label{eq:one-def-xi}
  \xi_A(z)=h_\gamma(z)V_\gamma(w,z)\psi_w\,,
\end{equation}
where $\psi_w$ is in $\ker(A^*-wI)$ and $V_\gamma(w,z)$ is given by
(\ref{eq:generalized-cayley}), then (\ref{eq:one-def-xi}) will satisfy
(P1) and (P2). Moreover, either by defining the involution as in
Theorem~\ref{thm:existence-of-commuting-involution} or by choosing $w$
and $\psi_w$ as in Theorem~\ref{thm:on-psi-w}, the function
(\ref{eq:one-def-xi}) also satisfies (P3). This follows either from
the proof of Theorem~\ref{thm:existence-of-commuting-involution} or
from the proof of Theorem~\ref{thm:on-psi-w} (see
\cite[Proposition~2.11]{us-4}). Note that, since $h_\gamma(z)$ is
defined up to a multiplying zero-free real entire function,
Lemma~\ref{lem:xi-up-to-function} implies that (\ref{eq:one-def-xi})
does not depend on the choice of the selfadjoint extension $A_\gamma$
nor on $w$ and, furthermore, every function $\xi_A:\C\to\cH$ can be
written as in (\ref{eq:one-def-xi}).

Fix $A\in\ournewclass$ and any function $\xi_A:\C\to\cH$ satisfying
(P1), (P2), and (P3). Then define
\begin{equation}
\label{eq:defining-phi}
\left(\Phi_A\varphi\right)(z):=\inner{\xi_A(\cc{z})}{\varphi},\qquad
	\varphi\in\cH.
\end{equation}
Due to (P1), $\Phi_A$ maps $\cH$ onto a certain linear manifold
$\Phi_A\cH$ of entire functions. The notation
$\widehat{\cH}=\Phi_A\cH$ will be used when it is no need of referring
to $A$. Note that if one fixes $z\in\C$ and allows $\varphi$ to run
over $\cH$, the inner product in (\ref{eq:defining-phi}) becomes a bounded
linear functional whose kernel is $\ran(A-zI)$. Hence, the complete
nonselfadjointness condition (\ref{eq:simplicity}) and the analyticity
of the functions in $\widehat{\cH}$ imply that $\Phi_A$ is injective.  A
generic element of $\widehat{\cH}$ will be denoted by
$\widehat{\varphi}(z)$, as a reminder of the fact that it is the image
under $\Phi_A$ of a unique element $\varphi\in\cH$. Clearly, the linear
space $\widehat{\cH}$ is turned into a Hilbert space by defining
\begin{equation*}
  \inner{\widehat{\eta}(\cdot)}{\widehat{\varphi}(\cdot)}:=
\inner{\eta}{\varphi}\,,
\end{equation*}
and $\Phi_A$ is an isometry from $\cH$ onto $\widehat{\cH}$.

\subsection{Properties of functional space}
\label{sec:prop-funct-space}
The properties of the isometry $\Phi_A$
and the space of functions $\widehat{\cH}$ previously defined are
discussed here.

The following assertion (see \cite{us-4}) follows from the properties
of the function $\xi_A(z)$ and the fact that
\begin{equation*}
  k(z,w):=\inner{\xi_A(\cc{z})}{\xi_A(\cc{w})}
\end{equation*}
is a reproducing kernel in  $\widehat{\cH}$
(cf. \cite[Proposition 1]{MR1821925}).
\begin{theorem}
  \label{thm:Phi-H-is-dB}
  Let $\Phi_A$ be defined by (\ref{eq:defining-phi}). For any operator
  $A\in\ournewclass$, the space of functions $\widehat{\cH}=\Phi_A\cH$
  is a de Branges space.
\end{theorem}

On the other hand, the isometry $\Phi_A$ transforms $A$ as expected:


\begin{theorem}
  \label{thm:func-model-properties}
  Fix an operator $A\in\ournewclass$. Let $J$ be the involution that
  appears in (P3) and $S$ be the operator of
  multiplication by the independent variable in the dB space
  $\widehat{\cH}=\Phi_A\cH$. Then, the following holds:
  \begin{enumerate}[(i)]
  \item $S=\Phi_AA\Phi_A^{-1}$ and $\dom(S)=\Phi_A\dom(A)$.
  \item $\#=\Phi_AJ\Phi_A^{-1}$.
  \item If $A_\gamma$ is a canonical selfadjoint extension of $A$,
    then $\Phi_AA_\gamma\Phi_A^{-1}$ is a canonical selfadjoint
    extension of $S$.
  \end{enumerate}
\end{theorem}


\subsection[Functional spaces for n-entire operators]
	       {Functional spaces for $\mb{\nentireclass{n}}$}
\label{sec:funct-spac-mbnent}
As already shown, to every operator $A\in\ournewclass$ there corresponds a 
dB space such that $A$ is unitarily equivalent to the operator $S$ of
multiplication by the independent variable in that dB space.
On the other hand, by (i) and (v) of
Theorem~\ref{thm:multiplication-operator}, the operator of
multiplication in every dB space $\mathcal{B}$ is an element of
$\mathscr{S}(\mathcal{B})$. The following assertion gives a
characterization of the dB spaces that correspond to operators in
$\nentireclass{n}$.
\begin{theorem}
  \label{thm:dB-spaces-for-En}
  Let $\Phi_A$ be defined by (\ref{eq:defining-phi}), with
  $A\in\ournewclass$, and $\widehat{\cH}=\Phi_A\cH$. For any
  $n\in\Z$, the operator $A$ is
  in $\nentireclass{n}$ if and only if $\assoc_n(\widehat{\cH})$
  contains a zero-free entire function.
\end{theorem}
This theorem follows directly from
Definitions~\ref{def:n-entireclass+} and \ref{def:n-entireclass-}, and
the properties of the $\xi_A(z)$, taking into account
(\ref{eq:space-decomposition}) (cf. \cite[Proposition
3.1]{us-4}).

In view of Theorem~\ref{thm:1-in-assoc-n-boosted} and
Remark~\ref{rem:zero-free-implies-real-zero-free}, one has the
following criterion for an operator to be in $\nentireclass{n}$
(cf. \cite[Proposition 3.7]{us-4}).
\begin{theorem}
  \label{thm:entire-woracek}
  Let $A_1$, $A_2$ be two canonical selfadjoint
  extensions of $A\in\ournewclass$. For any $n\in\Z$, the operator $A$
  is in $\nentireclass{n}$ if and only if the sequences $\spec(A_1)$
  and $\spec(A_2)$ comply with the conditions (C1), (C2), and (C3) of
  Theorem~\ref{thm:1-in-assoc-n-boosted}.
\end{theorem}

There are other results concerning the properties of the operator
classes $\nentireclass{n}$, $n\in\Z$, which are obtained by means of
the functional model given in this section. For instance,
\cite[Proposition 3.11]{us-4} states that for the definition of the
class $\nentireclass{n}$, with $n\in\Z^+$, it is sufficient to require
that (\ref{eq:def-En-classes}) holds for all $z\in\C$ with the
exception of a finite set. A more involved assertion stemming from the
functional model is the following one due to Strauss
\cite[Propositions 9 and 10]{MR1821925}.
\begin{theorem}
  \label{thm:strauss}
  Let $A$ be an operator in $\ournewclass$.
  \begin{enumerate}[(i)]
  \item $A\in\nentireclass{1}\setminus\nentireclass{0}$ if and
    only if there is an extension $B\supset A$ with empty spectrum
    (the resolvent set is the whole complex plane).
  \item $A\in\nentireclass{0}$ if and only if $A^{-1}$ has a
    quasinilpotent extension $K$ (that is, $\spec(K)=\{0\}$) such that
    $\dom(K)=\cH$ and $\ran(K)=\dom(A)$.
  \end{enumerate}
\end{theorem}

\subsection{Krein's entire operators}
\label{sec:kreins-entire-operators}

This section concludes with an elaboration of the relation between the notions
of operators entire and entire in the generalized sense introduced by Krein,
and the classes $\nentireclass{n}$ of $n$-entire operators.

According to Krein's terminology \cite[Section 2]{MR0011170}, a vector
$\mu\in\cH$ is said to be a \emph{gauge} for a densely defined operator
$A\in\ournewclass$ whenever
\begin{equation}
  \label{eq:gauge-equation}
  \cH=\ran(A-zI)\dotplus\Span\{\mu\}
\end{equation}
for some complex number $z=z_0$. Given a gauge, the set
\begin{equation}
  \label{eq:bad-set}
  \{z\in\C: \text{(\ref{eq:gauge-equation}) fails to hold}\}
\end{equation}
has no finite accumulation points and, therefore, its cardinality is at
most infinite countably. Furthermore, depending on the choice of the
gauge $\mu$, the set (\ref{eq:bad-set}) could be placed inside $\R$
\cite[Lemma 2.1]{us-2} or be contained outside $\R$ (see
\cite[Theorem 8c]{MR0011533} or \cite[Theorem 2.2]{us-2}). Krein calls a
gauge entire if the set (\ref{eq:bad-set}) turns out to be empty and,
in this case, the operator $A$ is entire \cite[Section 1]{MR0012177}
(see also \cite[Chapter 2, Section 5]{MR1466698}). Thus, by comparing
Definition~\ref{def:n-entireclass+} with (\ref{eq:gauge-equation}),
one concludes that the densely defined operators in $\nentireclass{0}$
correspond to Krein's class of entire operators.

For an entire operator $A$ with {\em real} entire gauge $\mu$, Krein defined the 
mapping
\begin{equation}
\label{eq:krein-functional-model}
\varphi\mapsto\hat{\varphi}(z)
	:=\frac{\inner{V_\gamma(w,\cc{z})\psi_w}{\varphi}}
			{\inner{V_\gamma(w,\cc{z})\psi_w}{\mu}},\quad \varphi\in\cH,
\end{equation}
where $w\in\C$ and $\psi_w$ are suitable chosen. Comparing with the functional 
model outlined in this section, one sees that \eqref{eq:krein-functional-model}
corresponds to a specialized choice of the function $h_\gamma(z)$ in
\eqref{eq:one-def-xi}, namely,
\[
  h_\gamma(z)=\frac{1}{\inner{V_\gamma(w,\cc{z})\psi_w}{\mu}}.
\]
Furthermore, due to the coincidence of the models in this case,
Krein's assertion that the existence of an entire gauge implies the
existence of a real entire gauge \cite[Theorem 8]{MR0011533} is a
simple consequence of
Remark~\ref{rem:zero-free-implies-real-zero-free}.

\begin{remark}
  \label{rem:differ-from-krein}
  In fact, Krein considered the mapping
  (\ref{eq:krein-functional-model}) not only for densely defined
  operators in $\nentireclass{0}$ but for all densely defined
  operators in $\ournewclass$, where $\mu$ is then an appropriately
  chosen element of $\cH$ \cite[Chapter~1, Section~2]{MR1466698}. Thus,
  Krein's functional space is a dB space if and only if $A$ is a densely
  defined operator in $\nentireclass{0}$ and $\mu$ is an entire gauge.
\end{remark}

In addition to the entire operators, Krein considered the so-called entire
operators in the generalized sense \cite[Section 4]{MR0235424}, which
were later studied in \cite{MR0301545} and \cite[Chapter
6]{MR0487539}. To the end of defining these operators, first note that
(\ref{eq:space-decomposition}) and Definition~\ref{def:n-entireclass+}
imply that $A\in\nentireclass{0}$ whenever
\begin{equation}
\label{eq:gauge-equation-alt}
  \inner{\xi_A(\cc{z})}{\mu}\ne 0\quad\text{for all}\quad z\in\C,
\end{equation}
and some fixed $\mu\in\cH$ (which, as already pointed out, can also be 
assumed real).
Now, take a densely defined operator $A\in\ournewclass$ and define
\[
\cH_+ := \dom(A^*),\text{ equipped with graph norm}.
\]
Then $\cH_+$ is a Hilbert space. Its dual is
\[
\cH_- := \{\text{anti-linear functionals $\cH_+$-continuous on $\cH_+$}\}.
\]
Clearly, one has $\cH_+\subset\cH\subset\cH_-$ (for details on
triplets of this kind ---the so-called Gelfand triplets--- refer to
\cite{MR0222718}).

With this setup, $A$ is said to be entire in the generalized sense if
there exists $\mu\in\cH_-\setminus\cH$ such that, for all $z\in\C$,
one has (\ref{eq:gauge-equation-alt}) with the inner product replaced
by the duality bracket between $\cH_+$ and $\cH_-$. \cite[Section
5]{us-2}. Note that this definition makes sense because
$\xi_A(z)\in\cH_+$ for all $z\in\C$.  Moreover, one can prove the
following \cite[Proposition 5.1]{us-2},
\[
\ds{\assoc_1\hat{\cH}}=\{\hat{\eta}(z) \text{ entire}:
	\hat{\eta}(z)=\inner{\xi_A(\cc{z})}{\eta}\text{ for some }\eta\in\cH_-\}.
\]
In view of \eqref{eq:gauge-equation-alt}, $A\in\ournewclass$ is then
entire in the generalized sense as long as $\assoc_1\hat{\cH}$
contains a zero-free, entire function (which can also be chosen
real). Recalling Definition \ref{def:associated-functions}, this
amounts to saying that there are vectors $\mu_0,\mu_1\in\cH$ such that
\begin{equation}
\label{eq:op-entire-generalized}
  \cH=\ran(A-zI)\dot{+}\Span\{\mu_0+z\mu_1\},
\end{equation}
for every $z\in\C$.  All in all, a densely defined operator
$A\in\ournewclass$ is entire in the generalized sense of Krein if and
only if it belongs to the class $\nentireclass{1}$.

The use of triplet of spaces, to define the notion of entire operators
in the generalized sense, can be replicated to a certain extent for
$n$-entire operators.  In \cite[Section 4]{us-4}, given
$A\in\ournewclass$ densely defined and $n\in\N$, a Gelfand triplet
$\cH_{+n}\subset\cH\subset\cH_{-n}$ is constructed in such a way that
$\cH_{-n}\cong\assoc_n\hat{\cH}$ and $\xi_A(z)\in\cH_{+n}$ for all
$z\in\C$. Then $A$ is $n$-entire if and only if there exists
$\mu\in\cH_{-n}$ such that
\[
\inner{\xi_A(\cc{z})}{\mu}_n\ne 0
\]
for all $z\in\C$; here $\inner{\cdot}{\cdot}_n$ denotes the duality
bracket between $\cH_{+n}$ and $\cH_{-n}$.

Unlike the construction due to Krein, the one sketched above is a bit
convoluted, rendering this alternative definition more difficult to use.

\section{Applications}

Most of the classical applications of the theory of entire operators
are discussed in detail in Gorbachuks' monography
\cite[Chapter~3]{MR1466698}; the first example below is probably the
one most frequently used as an illustration of an entire operator. In
fact, it was the first example (see \cite[Section~4 A]{MR0012177})
and, as mentioned in the Introduction, it is the germ of the
theory. For the definition of the operator, the exposition of this
first example follows \cite[Chapter~3 Section~1]{MR1466698} and
\cite[Section~1]{MR1627806}. The second example is also classical but
presented in a somewhat novel approach. Finally, the last example
illustrates a non trivial class of Schr\"odinger operators that are
$n$-entire with $n\in\N$ fixed but arbitrary (pedantically, one should
say that these Schr\"odinger operators are selfadjoint extensions of
some operator in $\nentireclass{n}$).

\subsection{The Hamburger moment problem}
\label{sec:hamburger}
Recall the formulation of this classical problem: Given a sequence
of real numbers $\{s_n\}_{n=0}^\infty$, one is interested in finding
conditions for the existence of a positive measure $m(x)$ such that
\begin{equation}
\label{eq:hamburger}
s_n = \int_{-\infty}^{\infty}x^n dm(x),\quad n\in\N\cup\{0\}.
\end{equation}
Assuming the existence of such a measure, one may ask whether it is
unique. If it is not unique, one is interested in describing all the
solutions to problem \eqref{eq:hamburger}. The moment problem is said
to be determinate when it has only one solution and indeterminate
otherwise. A very complete treatment of this classical problem can be
found in \cite{MR0184042}.

As it is well known, a necessary and sufficient condition for the
existence of a solution of \eqref{eq:hamburger} is that
\begin{equation}
\label{eq:positivity-hamburger}
\sum_{j,k=0}^n s_{k+j}\cc{z}_jz_k\ge 0,
\end{equation}
for every $n\in\N\cup\{0\}$ and arbitrary numbers $z_j\in\C$
\cite[Proposition~1.3]{MR1627806}.

Under the condition that \eqref{eq:positivity-hamburger} holds, one
considers the set $\cL$ of all polynomials in $\R$ with complex
coefficients,
\[
p(x)= \sum_{k=0}^n z_k x^k,\quad z_k\in\C,\quad n\in\N\cup\{0\},
\]
equipped with the sesquilinear form
\begin{equation}
\label{eq:inner-product-hamburger}
\inner{p}{q}:=\sum_{j=0}^n\sum_{k=0}^m s_{j+k}\cc{z}_jz_k.
\end{equation}
Then one obtains a Hilbert space $\cH$ as the completion of $\cL/\cL_0$ under
\eqref{eq:inner-product-hamburger}, where
\[
\cL_0:=\{p(x)\in\cL:\inner{p}{p}=0\}.
\]
In $\cH$ one defines the operator $A$, with domain $\cL/\cL_0$, as the lifting
of the operator defined by the mapping $p(x)\mapsto xp(x)$ with domain $\cL$.
This operator is symmetric and real with respect to (lifting to $\cL/\cL_0$ of)
the usual complex conjugation in $\cL$.

\begin{theorem}
\label{thm:determinate-indeterminate}
  Assume \eqref{eq:positivity-hamburger}. Let $A$ be the operator
  defined as above.  Then, either
\begin{enumerate}[(i)]
\item $A$ is essentially selfadjoint,
	in which case the problem \eqref{eq:hamburger} has a unique solution; or
\item the closure of $A$ has deficiency indices $(1,1)$, in which
	case the solution of \eqref{eq:hamburger} is not unique.
\end{enumerate}
\end{theorem}

It turns out that there is an orthonormal basis
$\{P_{k-1}(x)\}_{k=1}^\infty$ in $\cH$ such that $A$ has a Jacobi matrix
as its matrix representation with respect to it (see
\cite[Section~47]{MR1255973} for a discussion on the matrix
representation of unbounded symmetric operators). Note that one could
have taken a Jacobi matrix as the starting point for defining the
operator $A$ (see \cite[Chapter 4]{MR0184042}).

The element $P_k(x)$ of the basis mentioned above is a polynomial of
degree $k$ and it is known as the $k$-th orthogonal polynomial of the
first kind associated with the Jacobi matrix. It happens that
$P_0(x)\equiv 1$.

One has the following assertion \cite[Chapter~3,
Theorem~1.2]{MR1466698}.

\begin{theorem}
  \label{thm:jacobi-entire}
  If (ii) of Theorem~\ref{thm:determinate-indeterminate} takes place,
  then $P_0(x)$ is an entire gauge and, therefore,
  $A\in\nentireclass{0}$.
\end{theorem}

As a matter of fact $A$ is in $\nentireclass{-\infty}$. Indeed, as it
is straightforward to verify, $P_0(x)$ is in the domain of $A^n$ for
any $n\in\mathbb{N}$.


\subsection{The linear momentum operator}
\label{sec:linear-momentum}

In $\cH=L^2[-a,a]$, $0<a<+\infty$, consider the operator
\begin{equation*}
\dom(A)=\{\varphi(x)\in\text{AC}[-a,a]:\varphi(a)=0=\varphi(-a)\},
\quad A:=i\frac{d}{dx}.
\end{equation*}
Clearly, $A$ is closed and symmetric. Moreover,
\begin{equation*}
\dom(A^*) = \text{AC}[-a,a],\quad A^*=i\frac{d}{dx},
\end{equation*}
from which it is straightforward to verify that the deficiency indices
of $A$ are $(1,1)$.  The canonical selfadjoint extensions of $A$ can
be parametrized as
\begin{equation*}
\dom(A_\gamma) = \{\varphi(x)\in\text{AC}[-a,a]:
			\varphi(a)=e^{-i2\gamma}\varphi(-a)\},
\quad A_\gamma = i\frac{d}{dx},
\end{equation*}
for $\gamma\in{[}0,\pi)$. These selfadjoint extensions correspond to
different realizations of the linear momentum operator within the
interval $[-a,a]$. By a straightforward calculation,
\begin{equation}
\label{eq:spectrum-linear-momentum}
\Sp(A_\gamma)=\left\{\frac{\gamma+k\pi}{a}:k\in\Z\right\}.
\end{equation}
Clearly, the spectra are interlaced and their union equals $\R$ so it
follows that $A$ is regular, hence completely nonselfadjoint.

This operator can be shown to be entire in the generalized sense (that
is, 1-entire) by methods of directing functionals. This is the
classical approach discussed for instance in \cite{MR1466698}.  An
alternative method for showing that $A$ is in $\nentireclass{1}$ is
used here. This method resorts directly to Definition
\ref{def:n-entireclass+} and may be generalized to other differential
operators. Another example treated in a similar manner is found in
\cite[Example 3.4]{us-4}.

Define $\xi(x,z):=e^{-izx}$, $x\in [-a,a]$, $z\in\C$. This
zero-free entire function belongs to $\ker(A^* - zI)$ for all
$z\in\C$. For proving that $A$
is 1-entire it suffices to find $\mu_0(x),\mu_1(x)\in L^2[-a,a]$
such that
\begin{equation}
\label{eq:nice-example}
\int_{-a}^{a}e^{-iyx}\mu_0(x)dx + y \int_{-a}^{a}e^{-iyx}\mu_1(x)dx =1
\end{equation}
for all $y\in\R$ (and then use analytic continuation to the whole
complex plane). The search will be guided by formally taking the
inverse Fourier transform of \eqref{eq:nice-example} and switching
without much questioning the order of integration, obtaining in that
way the differential equation
\[
\mu_0(x) - i\mu'_1(x) = \delta(x),
\]
where $\delta(x)$ is the Dirac's distribution. This equation suggests to set
\begin{align}
\mu_0(x) &= \frac{1}{2a}\chi_{[-a,a]}(x)\label{eq:mu-0}
\\[1mm]
\mu_1(x) &= - i\frac{a+x}{2a}\chi_{[-a,0]}(x)
		    + i\frac{a-x}{2a}\chi_{[0,a]}(x),\label{eq:mu-1}
\end{align}
where $\chi_S(x)$ denotes the characteristic function of the set $S$.
A simple computation shows that indeed \eqref{eq:mu-0} and
\eqref{eq:mu-1} satisfy \eqref{eq:nice-example} thus $A$ is
1-entire as asserted.

This operator is associated with a Paley-Wiener space. Indeed, it
is apparent that
\[
\cPW_a = \left\{\hat{\varphi}(z) 
				=\int_{-a}^a\xi(x,z)\varphi(x) dx:
				 \varphi(x)\in L^2[-a,a] \right\};
\] 
notice that here one has an instance of application of the abstract 
functional model discussed in this work. Also, notice that this implies
the sharper statement $A\in\nentireclass{1}\setminus\nentireclass{0}$.

\subsection{Spectral analysis of radial Schr{\"o}dinger operators}
\label{sec:radial-schrodinger-operator}

Consider the selfadjoint operators that arise from the differential
expression
\[
\tau:=
	-\frac{d^2}{dx^2} + \frac{l(l+1)}{x^2} + q(x),
	\quad x\in(0,1),\quad  l\ge-\frac12,
\]
along with separated selfadjoint boundary conditions.
These operators describe the radial part of the Schr\"odinger operator
for a particle confined to a ball of finite radius, when the potential is
spherically symmetric.

The potential function $q(x)$ is assumed real such that
$\tilde{q}(x)\in L_1(0,1)$, where
\[
\tilde{q}(x):=\left\{\begin{array}{ll}
					 xq(x) & l>-\tfrac12,
					 \\[1mm]
					 x(1-\log x)q(x) & l=-\tfrac12.
			         \end{array}
			   \right.
\]
Under this hypothesis, it is shown in
\cite[Theorem 2.4]{kostenko-1} that $\tau$ is regular
at $x=1$, and the limit point case (resp. limit circle case) at $x=0$ for
$l\ge 1/2$ (resp. $l\in[-1/2,1/2)$). If $\tau$ is in the limit circle
case at $x=0$,  it is usual to
add the boundary condition
\begin{equation}
\label{eq:boundary-condition-at-0}
\lim_{x\to 0}x^l\left[(l+1)\varphi(x) - x\varphi'(x)\right]=0.
\end{equation}
Other boundary conditions can serve as well. A comprehensive investigation of
them can be found in \cite{bulla}.

With this setup $\tau$ gives rise to a family of
selfadjoint operators $H_\beta$, for $\beta\in[0,\pi)$, associated to the
boundary conditions $\varphi(1)\cos\beta = \varphi'(1)\sin\beta$.
These operators are the
canonical selfadjoint extensions of a certain closed, regular, symmetric
operator $H$, having deficiency indices $(1,1)$. In
\cite[Theorem 4.3 and Corollary 4.4]{us-6}, the following statement is
proven:

\begin{theorem}
Let $l\ge -\frac12$ and assume that $\tilde{q}(x)$ belongs to $L_p(0,1)$, with
$p>2$. Then,
\begin{enumerate}[(i)]
\item the operator $H$ is $n$-entire if and only if $n>\frac{l}{2}+\frac34$.
	  In that case,
\item the spectra of two canonical selfadjoint extensions
  	  $H_{\beta_1}$, $H_{\beta_2}$ of $H$ satisfy conditions (C1),
  	  (C2) and (C3).
\end{enumerate}
\end{theorem}
\def\cprime{$'$} \def\lfhook#1{\setbox0=\hbox{#1}{\ooalign{\hidewidth
  \lower1.5ex\hbox{'}\hidewidth\crcr\unhbox0}}}

\end{document}